\ifx\pdfoutput\undefined
	\documentclass[aps,prd,a4paper,twoside,twocolumn,
	noshowkeys,nofootinbib,amssymb,amsmath,amsfonts,showpacs,dvips]{revtex4}
\else
	\documentclass[aps,prd,a4paper,twoside,twocolumn,superscriptaddress,
	noshowkeys,nofootinbib,amssymb,amsmath,amsfonts,showpacs,pdftex]{revtex4}
\fi

\usepackage{amsmath}
\usepackage{amssymb}
\usepackage{graphicx}
\usepackage{dcolumn}

\newcommand{\geff}{g^\mathrm{eff}}
\newcommand{\Neff}{N^\mathrm{eff}}
\newcommand{\DNoth}{\Delta\Neff_{\mathrm{others}}}
\newcommand{\conc}{\Lambda\mathrm{CDM}}
\newcommand{\params}{\omega_b,\, \omega_m,\, n,\, \tau,\, A,\, \omega_\nu,\, \xi, \DNoth}
\newcommand{\coreparams}{\omega_b,\,\omega_m,\,n,\, \tau,\, A}

\newcommand{\apjs}{Astrophys. J. Suppl. Ser.}
\newcommand{\nphysb}{Nucl. Phys. B}
\newcommand{\aaph}{Astron. \& Astrophys.}
\newcommand{\plb}{Phys. Lett. B}
\newcommand{\mnras}{Mon. Not. R. Astron. Soc.}
\newcommand{\cqg}{Class. Quantum Grav.}

\DeclareMathOperator{\eV}{eV}

\begin{document}

\title{Joint constraints on the lepton asymmetry of the Universe and neutrino mass from the Wilkinson Microwave Anisotropy Probe}
\author{Massimiliano Lattanzi}
\email{lattanzi@icra.it}
\author{Remo Ruffini}
\author{Gregory V. Vereshchagin}
\affiliation{ICRA --- International Center for Relativistic Astrophysics and \\ Dipartimento di Fisica, Universit\`a di Roma ``La Sapienza'', Piazzale Aldo Moro 2, I-00185 Roma, Italy.}
\begin{abstract}
We use the Wilkinson Microwave Anisotropy Probe (WMAP) data on the spectrum of cosmic microwave background anisotropies to put constraints on the present amount of lepton asymmetry $L$, parameterized by the dimensionless chemical potential (also called degeneracy parameter) $\xi$ and on the effective number of relativistic particle species. We assume a flat cosmological model with three thermally distributed neutrino species having all the same mass and chemical potential, plus an additional amount of effectively massless exotic particle species. The extra energy density associated to these species is parameterized through an effective number of additional species $\DNoth$. We find that $0<|\xi|<1.1 $ and correspondingly $0<|L|<0.9$ at $2\sigma$, so that WMAP data alone cannot firmly rule out scenarios with a large lepton number; moreover, a small preference for this kind of scenarios is actually found. We also discuss the effect of the asymmetry on the estimation of other parameters and in particular of the neutrino mass. In the case of perfect lepton symmetry, we obtain the standard results. When the amount of asymmetry is left free, we find $\sum m_\nu < 3.6$~eV at $2\sigma$. Finally we study how the determination of $|L|$ is affected by the assumptions on $\DNoth$. We find that lower values of the extra energy density allow for larger values of the lepton asymmetry, effectively ruling out, at $2\sigma$ level, lepton symmetric models with $\DNoth\simeq 0$.
\end{abstract}
\maketitle

\section{Introduction}
It is a remarkable fact that our observational knowledge of the Universe can be justified in terms of a model, the so-called power-law $\conc$ model, characterized by just six parameters, describing the matter content of our Universe (the physical density of baryons $\omega_b$, the physical density of matter $\omega_m$, the Hubble constant $h$), the initial conditions from which it evolved (the amplitude $A$ and the spectral index $n$ of the primordial power spectrum) and the optical depth at reionization $(\tau)$. In particular, this model provides a good fit to both the cosmic microwave background (CMB) \cite{WMAP:parameters} and large scale structure (LSS) data (although in this last case one additional parameter, the bias parameter $b$, is needed) \cite{Te04}. Nevertheless, the data leave room for more refined models, described by additional parameters: among them, the spatial curvature, the amplitude of tensor fluctuations, a running spectral index for scalar modes, the equation of state for dark energy, the neutrino fraction in the dark matter component, a non-standard value for the relativistic energy density. All have been considered in previous works. In particular the last two have been studied in order to gain deeper information on the properties of neutrinos \cite{Ha02, WMAP:parameters, Ha03, El03, El02, Allen03, Barger04, Cr03,Pi03, DiBari02, DiBari03, Barger03, Cr04, Ha04, Ha03b, Cu03}.
Before the measurements of the CMB anisotropy spectrum carried out by the Wilkinson Microwave Anisotropy Probe (WMAP) \cite{WMAP:mission, WMAP:Hinshaw, WMAP:Kogut, WMAP:Page, WMAP:Peiris, WMAP:parameters,WMAP:methodology}, the combined CMB and LSS data yielded the following upper bound on the sum of neutrino masses: $\sum m_\nu \le 3$~eV \cite{Ha02}. The WMAP precision data allowed to strengthen this limit. Using rather simplifying assumptions, i.e., assuming three thermalized neutrino families all with the same mass and a null chemical potential (thus implying perfect lepton symmetry), the WMAP team found that the neutrino mass should be lower than 0.23 eV \cite{WMAP:parameters}. This tight limit has been somewhat relaxed to $\sum m_\nu \le 1$~eV \cite{Ha03} owing to a more careful treatment of the Ly-$\alpha$ data, and its dependence on the priors has been examined \cite{El03}. The LSS data can also be used to put similar constraints, although they are usually weaker. Using the data from the 2 Degree Field Galaxy Redshift Survey (2dFGRS) and assuming ``concordance'' values for the matter density $\Omega_m$ and the Hubble constant $h$ it is found that $\sum m_\nu \le 1.8$~eV \cite{El02}. A combined analysis of the Sloan Digital Sky Survey (SDSS) and WMAP data gives a similar bound : $\sum m_\nu \le 1.7$~eV \cite{Te04}. Quite interestingly, the authors of Ref. \cite{Te04} claim that, from a conservative point of view (i.e., making as few assumptions as possible), the WMAP data alone don't give any information about the neutrino mass and are indeed consistent with neutrinos making up the 100\% of dark matter. In Ref. \cite{Allen03} it is claimed that the cosmological data favor a non-zero neutrino mass at the 68\% confidence limit, while the authors of Ref. \cite{Barger04} find the limit $\sum m_\nu<0.74$.

At the same time, more detailed scenarios with a different structure of the neutrino sector have been studied. The first and more natural extension to the standard scenario is the one in which a certain degree of lepton asymmetry (parameterized by the so-called \emph{degeneracy parameter} $\xi$, i.e., the dimensionless chemical potential) is introduced \cite{Freese83, Ruffini83, Ruffini88}. Although standard models of baryogenesis (for example those based on $SU(5)$ grand unification models) predict the lepton charge asymmetry to be of the same order of the baryonic asymmetry $B\sim 10^{-10}$, nevertheless there are many particle physics motivated scenario in which a lepton asymmetry much larger than the baryionic one is generated \cite {Harvey81, Dolgov91, Dolgov92,Foot96, Casas99, March99, Dolgov00, McDonald00, Kawasaki02, DiBari02b, Yama03, Chiba04, Taka04}. In some cases, the predicted lepton asymmetry can be of order unity. One of the interesting cosmological implications of a net leptonic asymmetry is the possibility to generate small observed baryonic asymmetry of the Universe \cite{Buch04,Falc01} via the so-called sphaleron process \cite{Kuz85}. The process of Big Bang Nucleosynthesis (BBN) is very sensitive to a lepton asymmetry in the electronic sector, since an excess (deficit) of electron neutrinos with respect to their antiparticles, alters the equilibrium of beta reactions and leads to a lower (higher) cosmological neutron to proton ratio $n/p$. On the other hand, an asymmetry in the $\mu$ or $\tau$ sector, even if not influencing directly the beta reactions, can increase the equilibrium $n/p$ ratio due to a faster cosmological expansion. This can be used to constrain the value of the degeneracy parameter \cite{Bi91}. This leads to the bounds $-0.01 <\xi_e<0.22$ and $|\xi_{\mu,\tau}|<2.6$ \cite{Kn01,Hans02}.

The effect of a relic neutrino asymmetry on the CMB anisotropy and matter power spectrum was first studied in Ref. \cite{Le99}, and is mainly related to the fact that a lepton asymmetry implies an energy density in relativistic particles larger than in the standard case. The cosmological observables can then be used to constrain this extra energy density, parameterized by the effective number of relativistic neutrino species $\Neff$. Although this is somewhat more general than the case of a lepton asymmetry, in the sense that the extra energy density can arise due to other effects as well, nevertheless the case of a non-null chemical potential is not strictly covered by the introduction of $\Neff$. This is because the increased relativistic energy density is not the only effect connected to the lepton asymmetry (an additional side effect is for example a change in the Jeans mass of neutrinos \cite{Freese83, Ruffini86, Ruffini88}). In the hypothesis of a negligible neutrino mass, it has been shown that the WMAP data constrain $\Neff$ to be smaller than 9; when other CMB and LSS data are taken into account, the bound shrinks to $1.4 \le \Neff \le 6.8$ \cite{Cr03,Pi03}. A combined analysis of CMB and BBN data leads to even tighter bounds \cite{DiBari02,DiBari03,Ha03,Barger03}. A more detailed analysis, in which the effective number of relativistic relics and the neutrino mass are both left arbitrary and varied independently, can be found in Ref. \cite{Cr04}. In the same paper, the effect of different mass splittings is also studied. Finally an extension of these arguments to the case in which additional relativistic, low-mass relics (such as a fourth, sterile neutrino or a QCD axion) are present, has been studied in Ref. \cite{Ha04}. 

The goal of this paper is to perform an analysis of the WMAP data using the degeneracy parameter, together with the effective number of relativistic particles, as additional free parameters, in order to put constraints on the lepton number of the Universe. We work in the framework of an extended cosmological model with three thermally distributed neutrino families having all the same mass and chemical potential, plus a certain amount of exotic particles species, considered to be effectively massless.  We use the physical neutrino density $\omega_\nu\equiv\Omega_\nu h^2$, the degeneracy parameter $\xi$ and the extra energy density in exotic particles $\DNoth$ as additional parameters that describe the neutrino sector. We perform an analysis in a 8-dimensional parameter space that includes the standard, ``core'' cosmological parameters. 

The paper is organized as follows. After a discussion on the motivations that drive our work in Sec. \ref{sec:Motivation}, we shortly review some basic formulae in Sec. \ref{sec:Basic formulae} and discuss the impact of a non-null degeneracy parameter on the CMB spectrum in Sec. \ref{sec:Effect}. In Sec. \ref{sec:Method} we describe the analysis pipeline, while in Sec. \ref{sec:Results} we present our basic results. Finally, we draw our conclusions in Sec. \ref{sec:Conclusions}.

\section{Motivation for this work}
\label{sec:Motivation}

The main motivation for this work comes from the fact that, even though several analyses have been performed which were aimed at putting constraints on the number of effective relativistic degrees of freedom, a statistical analysis of the CMB data aiming to put bounds directly on the degeneracy parameter, instead of on $\Neff$, is nevertheless still missing. There are two reasons for this: first of all, in the limit of a vanishing neutrino mass, the increase in $\Neff$ is in effect all that is needed to implement the non null chemical potential into the standard model of the evolution of perturbations \cite{Le99, Ma95}. It is then argued that, since neutrinos with mass smaller than roughly $0.3$~eV, being still relativistic at the time of last scattering, would behave as massless, the distinction between $\xi$ and $\Neff$ is no more relevant in this case for what concerns their effect on the CMB anisotropy spectrum. Although this is certainly true, it is our opinion that this does not allow to neglect \emph{a priori} the difference between the two parameters. One reason is that the most conservative bound on neutrino mass, coming from the tritium beta decay experiments, reads $m_\nu < 2.2$~eV (at the $2\sigma$ level) \cite{Wein99,Bonn02}, that is quite higher than the value of 0.3~eV quoted above. The main evidences for a neutrino mass in the sub-eV range come, in the field of particle physics, from the experiments on neutrinoless double beta decay \cite{Kl02, Kl04}, whose interpretation depends on assumptions about the Majorana nature of neutrinos and on the details of the mixing matrix. Other indications of a sub-eV mass come, as stated above, from cosmology and in particular from the power spectrum of anisotropies, but since we want to keep our results as much as possible independent from other analyses, we should not use information on neutrino mass derived from the previous analyses of the CMB data.
Moreover, let us note that CMB data analyses are often refined using the results from LSS experiments. Since the structure formation, starting close to the epoch of matter-radiation equality, goes on until very late times, even very light neutrinos (in the range $10^{-3}\div0.3$~eV) cannot be considered  massless for the purpose of evaluating their effect on the matter power spectrum. This means in particular that using $\Neff$ would lead to overlook the change in the free streaming length and in the Jeans mass of neutrinos due to the increased velocity dispersion {\cite{La03}}. It is then our opinion that the use of $\Neff$, even if correct with respect to the interpretation of CMB data, precludes the possibility of correctly implementing the LSS data as a subsequent step in the analysis pipeline. 

The second point against the cosmological significance of the degeneracy parameter is related to the constrain from BBN. It was recently shown that, if the Large Mixing Angle (LMA) solution to the solar neutrino problem is correct (as the results of the KamLAND experiment suggest \cite{KamLAND03}), then the flavor neutrino oscillations equalize the chemical potentials of $e$, $\mu$ and $\tau$ neutrinos prior to the onset of BBN, so that a stringent limit $\xi\lesssim 0.07$ actually applies to all flavours \cite{Do02, Ab02, Wo02}. This would constrain the lepton asymmetry of the Universe to such small values that it could be safely ignored in cosmological analyses. However, the presence of another relativistic particle or scalar field would make these limits relax \cite{Barger03b}, while the effect of the mixing with a light sterile neutrino, whose existence is required in order to account for the results of the Liquid Scintillation Neutrino Detector (LSND) experiment \cite{LSND}, is still unclear \cite{Ab02}. Moreover, it has been recently shown that a hypothetical neutrino-majoron coupling can suppress neutrino flavor oscillations, thus reopening a window for a large lepton asymmetry \cite{Do04, Do04b}. For all these reasons, we judge it is interesting to study if CMB data alone can constraint or maybe even rule out such exotic scenarios.

\section{Basic formulae}
\label{sec:Basic formulae}

It is customary in cosmology to call ultrarelativistic (or simply relativistic) a species $x$ that decouples from the photon bath at a temperature $T_d$ such that its thermal energy is much larger than its rest mass energy: $k_B T_d \gg m_x c^2$.

Owing to Liouville's theorem, the distribution function in momentum space $f_x(p\,;\,T_x,\xi_x)$ of the species $x$ is given, after decoupling, by (we shall use all throughout the paper units in which $c=\hbar=k_B=1$):
\begin{equation}
f_x(p\,;\,T_x,\xi_x) =\frac{g_x}{(2\pi)^3}\left[\exp\left(\frac{p}{T_x}-\xi_x\right)\pm1\right]^{-1},
\end{equation}
where $\xi \equiv \mu_d/T_d$ is the dimensionless chemical potential, often called degeneracy parameter, the sign $+\,(-)$ corresponds to the case in which the $x$'s are fermions (bosons), $g$ is the number of quantum degrees of freedom, and the temperature $T$ evolves in time as the inverse of the cosmological scale factor $a$, so that $T(t)\cdot a(t)=\mathrm{const.}$

The energy density of the $x$'s at a given temperature is readily calculated:
\begin{multline}
\rho_x(T_x,\xi)=\int \,E(p)\,f(p\,;\,T_x,\xi_x) d^3{\vec p}=\\[0.2cm]
=\frac{g_x}{2\pi^2}\int_0^\infty \,p^2 \sqrt{p^2+m_x^2}\,f(p\,;\,T_x,\xi_x)dp.
\label{eq:rhox}
\end{multline}
Using the dimensionless quantities $y\equiv p/T$ and $\beta\equiv~m_x/T$, the expression for the energy density can be put in the form:
\begin{equation}
\rho_x(T_x,\xi_x)
=\frac{g_x}{2\pi^2}T_x^4\int_0^\infty dy\,y^2 \frac{\sqrt{y^2+\beta^2}}{\exp(y-\xi_x)\pm 1}.
\end{equation}
We stress the fact that a temperature dependence is still present in the integral through the term $\beta$. However, the temperature dependence disappear from the integral in two notable limits, the ultrarelativistic (UR) and non-relativistic (NR) one, corresponding respectively to the two opposite cases $\beta\ll 1$ and $\beta\gg 1$ \cite{Ruffini83}. Then, defining 
\begin{equation}
J_n^\pm(\xi) \equiv \left( \int_0^\infty \frac{y^n}{e^{y-\xi}\pm 1}dy \right)
\left( \int_0^\infty \frac{y^n}{e^{y}\pm 1}dy \right)^{-1},
\end{equation}
so that $J_n^\pm(0)=1$, we have
\begin{equation}
\rho_x(T_x,\xi_x)=\left\{
\begin{array}{l}
\left(
\begin{array}{c}
1 \\
7/8
\end{array}
\right)
g_x \displaystyle \frac{\pi^2}{30} J_3^\pm(\xi_x) T_x^4 \qquad \mathrm{UR} \\[0.5 cm]
\left(
\begin{array}{c}
1 \\
3/4
\end{array}
\right)
g_x \displaystyle \frac{\zeta(3)}{\pi^2} m_x J_2^\pm(\xi_x) T_x^3  \qquad \mathrm{NR}
\end{array}
\right.
\end{equation}
where the upper and lower values in parentheses in front of the expression in the right-hand side hold for bosons and fermions respectively, and $\zeta(n)$ is the Riemann Zeta function of order $n$.

It is useful to express $\rho_x(t)$ in terms of the present day energy density of the cosmic background photons:
\begin{multline}
\rho_x(t) 
=
\left(
\begin{array}{c}
1 \\
7/8
\end{array}
\right)
 \left[ \frac{g_x}{2} \left( \frac{T^0_x}{T^0_\gamma}
\right)^4 J_3^\pm(\xi_x) \right] \rho_\gamma^0 (1+z)^4
\equiv\\[0.2cm]
\equiv g_x^\mathrm{eff}\rho_\gamma^0 \left(1+z\right)^4,
\label{eq:rhox_UR}
\end{multline}
having defined an effective number of relativistic degrees of freedom $g_x^\mathrm{eff}$ as
\begin{equation}
g_x^{\mathrm{eff}}
\equiv
\frac{g_x}{2}
\left(
\begin{array}{c}
1 \\
7/8
\end{array}
\right)
\cdot
 \left[ \left( \frac{T^0_x}{T^0_\gamma}
\right)^4 J_3^\pm(\xi_x) \right].
\label{eq:geff_def}
\end{equation}
It is often the case that one has to consider a fermion species $x$ together with its antiparticle $\bar x$, the most notable example being the relic neutrinos and antineutrinos.
In chemical equilibrium, the relation $\xi_x = -\xi_{\bar x}$ holds owing to the conservation of chemical potential, as can be seen considering the reaction:
\begin{equation}
x+\bar x \longleftrightarrow\,\ldots\,\longleftrightarrow \gamma+\gamma
\end{equation}
and noting that the chemical potential in the final state vanishes \cite{book:Weinberg}. This relation holds for neutrinos and antineutrinos in several cosmological scenarios. There are some exceptions to this, most notably early Universe scenarios in which lepton asymmetry is generated \cite{Fo97} or destroyed \cite{Ab04} by active-sterile neutrino oscillations at low temperatures. However, we shall assume all throughout the paper that the relation $\xi_\nu = -\xi_{\bar \nu}$ holds.

It can then be shown that
\begin{multline} 
\geff_{x+\bar x} = 
\geff_x + \geff_{\bar x} = \\[0.2cm]
= \frac{7}{8}g_x \left[1 + \frac{30}{7} \left( \frac{\xi_x}{\pi} \right)^2 + \frac{15}{7} \left( \frac{\xi_x}{\pi} \right)^4 \right] \left( \frac{T^0_x}{T^0_\gamma}
\right)^4,
\label{eq:geff x+xbar}
\end{multline}
where the factor between square parentheses can be recognized as what it is often quoted as the contribution of a non-vanishing chemical potential to the effective number of relativistic species $\Neff$.

The definitions introduced above can be easily extended to the case when several ultra-relativistic species $x_i$ are present:
\begin{equation}
\geff
\equiv
\sum_i
\geff_i,
\end{equation}
where photons are excluded from the summation. This means that, since $\geff_\gamma=g_\gamma=1$, the actual number of relativistic degrees of freedom is $(1+\geff)$.

The total density of ultrarelativistic particles  at a given time is thus:
\begin{equation}
\rho_\mathrm{rad}=\rho_\gamma^0\left(1+\geff\right)(1+z)^4.
\end{equation}

Finally we can use this expression to find the dependence on $\geff$ of the redshift of radiation-matter equality $z_{eq}$ (the subscripts $b$ and $\mathrm{CDM}$ stands for baryons and cold dark matter respectively):

\begin{equation}
1+z_{eq}=\frac{\rho_b^0+\rho_{\mathrm{CDM}}^0}{\rho_\gamma^0}
\left(1+\geff\right)^{-1}.
\end{equation}

So, the larger is the energy density of ultra-relativistic particles in the Universe, parameterized by the effective number of degrees of freedom $\geff$, the smaller $z_{eq}$ will be, i.e., the later the equality between radiation and matter will occur. In other words, supposing that the density in non-relativistic particles (baryons + CDM) is well known and fixed, having more relativistic degrees of freedom will shift $z_{eq}$ closer to us and to the CMB decoupling.

In the standard cosmological scenario, the only contribution to the energy density of relativistic particles other than photons is the one due to the three families of standard neutrinos, with zero chemical potential. The ratio of the neutrino temperature to the photon temperature is $T^0_\nu/T^0_\gamma=(4/11)^{1/3}$,  due to the entropy transfer that followed the electron-positron annihilation, shortly after neutrino decoupling. Then

\begin{equation}
\geff=\frac{7}{8}\left(\frac{4}{11}\right)^{4/3}N_{\nu}\simeq 0.23\,N_\nu ,
\label{eq:geff vs N}
\end{equation}

where $N_\nu=3$ is the number of neutrino families. The energy density in a single neutrino species is:

\begin{equation}
\rho_\nu^{\mathrm{std}}=\frac{7\pi^2}{120} \left(\frac{4}{11}\right)^{4/3}T_\gamma^4.
\end{equation}

However, several mechanisms that could increase (or even decrease) the energy density of relativistic particles have been proposed. In the presence of some extra relics (such as sterile neutrinos, majorons, axions, etc.) the energy density of radiation would obviously increase. A non-zero chemical potential for neutrinos or an unaccounted change of $\rho_\gamma$, due for example to particle decays that increase the photon temperature, would produce the same result. In all cases the effect is the same: a change in $\geff$, as it can be seen by looking at eq. (\ref{eq:geff_def}). It is usual in the literature to parameterize the extra energy density by introducing an effective number of neutrino families $\Neff$, defined as:

\begin{equation}
\Neff \equiv \frac{\sum_i \rho_i}{\rho_\nu^{\mathrm{std}}},
\end{equation}

where again the sum runs over all ultrarelativistic species with the exceptions of photons. It is clear from this definition that $\Neff$ is actually the energy density in ultrarelativistic species (apart from photons) normalized to the energy density of a single neutrino species with zero chemical potential and standard thermal history. It is easy to show that a relation formally similar to (\ref{eq:geff vs N}) holds in the non-standard scenario:

\begin{equation}
\geff=0.23\,\Neff.
\end{equation}

In addition, it should be noted that even in the standard scenario $\Neff\neq N_\nu=3$, but instead $\Neff\simeq 3.04$. This is due to the fact that neutrino decoupling is not instantaneous, so that neutrino actually share some of the entropy transfer of the $e^+ e^-$ annihilation, on one side, and to finite temperature quantum electrodynamics corrections on the other \cite{Dolgov97, Mangano02}.

It is also useful to introduce the effective number of additional relativistic species $\Delta\Neff$ defined as:
\begin{equation}
\Delta \Neff\equiv\Neff-3.04\,,
\end{equation}
so that $\Delta \Neff=0$ in the standard scenario. Please note that $\Delta \Neff$ can also be negative, for example in very low reheating scenarios \cite{Giud01}.

In this paper we shall consider a scenario in which the radiation content of the Universe at the time of radiation-matter equality is shared among photons, three neutrino families with standard temperature but possibly non-zero chemical potential, and some other relic particle. We shall suppose that the presence of the latter can be completely taken into account through its effect on $\Neff$. This is true if the species has been in its ultrarelativistic regime for the most part of the history of the Universe. The presence of this extra relic is required for our analysis, in order to circumvent the equalization of neutrino chemical potentials, as explained at the end of section \ref{sec:Motivation}. We also assume that the degeneracy parameters for neutrinos and antineutrinos are equal and opposite, and that $e$, $\mu$ and $\tau$ neutrinos all have the same chemical potential.

The extra energy density can thus be split into two distinct contributions, the first due to the non-zero degeneracy parameter of neutrinos and the second due to the extra relic(s): 
\begin{equation}
\Delta \Neff= \Delta\Neff_\nu(\xi)+\Delta\Neff_{\mathrm{others}}.
\label{eq:DNtot}
\end{equation}
Following our assumptions, $\Delta\Neff_\nu$ can be expressed as a function of the chemical potential only:
\begin{equation}
\Delta\Neff_\nu(\xi)=3\left[\frac{30}{7} \left( \frac{\xi_x}{\pi} \right)^2 + \frac{15}{7} \left( \frac{\xi_x}{\pi} \right)^4 \right].
\label{eq:DNnu}
\end{equation}

\section{Effect of a non-null chemical potential}
\label{sec:Effect}

As anticipated above, the main effect connected to the presence of a non-vanishing degeneracy parameter, is an increase in $\geff$ (or, equivalently, in $\Neff$). The presence of this extra number of effective relativistic degrees of freedom can in principle be detected from observations of the CMB radiation. The shift of matter-radiation equality has important consequences for the CMB anisotropy spectrum, these being due to the larger amplitude of the oscillations that enter the horizon during the radiation dominated phase, and to a larger early integrated Sachs-Wolfe (ISW) effect. However these effects, basically due to the speeding up of the cosmological expansion, can be similarly produced by the variation of other cosmological parameters, for example by a smaller CDM density.

Moreover since the change in the redshift of matter-radiation equality depends on the ultra-relativistic species only through the quantity $\geff$, it cannot be used to distinguish between the different species (i.e., it is ``flavor blind''), nor to understand if the excess energy density is due to the presence of some unconventional relic, to an extra entropy transfer to photons, or to a non-null chemical potential (i.e., to a lepton asymmetry), or maybe to all of the previous.

However ultrarelativistic particles, other than changing the background evolution, have an effect even on the evolution of perturbations, as it was pointed out in \cite{Ba04} with particular regard to the case of neutrinos. First of all, the high velocity dispersion of ultrarelativistic particles damps all perturbations under the horizon scale. Second, the anisotropic part of the neutrino stress-energy tensor couples with the tensor part of the metric perturbations. It was shown in \cite{We04} that this reduces the amplitude squared of tensor modes by roughly 30\% at small scales. Finally, the authors of Ref. \cite{Ba04} claimed that the perturbations of relativistic neutrinos produce a distinctive phase shift of the CMB acoustic oscillations. These effects can thus be used to break the degeneracy between $\geff$ and other parameters. It remains to establish whether they can be used to break the degeneracy between the different contributions to $\geff$ or not. Even without performing a detailed analysis, it can be seen by looking at the relevant equations in Ref. \cite{Ba04} and \cite{We04}, that both the absorption of tensor modes and the phase shift depend on the quantity $f_\nu \equiv \rho_\nu/(\rho_\gamma + \rho_\nu)$. The effect of free streaming, even if more difficult to express analytically, is also mainly dependent on the value of $f_\nu$ \cite{Hu98}. If we consider the case where the three standard neutrinos are the only contribution to the radiation energy density other than photons, but we allow for the possibility of a non-vanishing chemical potential or for a different $T_\nu^0/T_\gamma^0$ ratio, we see again that the changes in the shape of the CMB anisotropy spectrum depend only on $\geff$ as whole, as long as eq. (\ref{eq:rhox_UR}) retains its validity, i.e., as long as neutrinos are in their ultra-relativistic regime.
Even considering the presence of some additional relic particle $x$ does not seem to change this picture. Supposing that the other ultrarelativistic particles behave as neutrinos for what concerns the effects under consideration, we can argue that in all cases the relevant quantity is $f_x\equiv\rho_x/\rho_\mathrm{rad}$, so that we are again lead to the conclusions that $\geff$ is the only relevant parameter. This means, for example, that in the case of neutrinos with mass less than $\sim 0.3$~eV, so that they stay ultrarelativistic until the time of last scattering and for some time after, the effect on CMB perturbations is exactly the same of massless neutrinos, and every change in their temperature or chemical potential, as even the presence of an additional, sterile neutrino, is absorbed in $\geff$ (moreover, we don't have obviously any possibility to extract information about their mass).

However the picture changes when considering neutrinos (or other relic particles) that go out from the ultrarelativistic regime before matter-radiation decoupling. If the neutrino mass is larger than $\sim 0.3$~eV, the effect of its finite mass is felt by the perturbations that enter the horizon after neutrinos have gone out from the ultrarelativistic regime, because from some point on the evolution the energy density will be given no more by the approximate formula (\ref{eq:rhox_UR}), but instead by eq. (\ref{eq:rhox}) that contains a dependence on mass through the term $\beta$. A side effect of this is that it will not be possible to single out the dependence of $\rho$ from $T$ and $\xi$ as an overall factor, so that these two contributions become distinguishable.

Let us make this more clear with an example. Consider a gravitational wave entering the horizon after neutrinos became non-relativistic, but before matter-radiation decoupling. This wave will be absorbed, according to \cite{We04}, proportionally to $\rho_\nu$. On the other hand the free streaming length of neutrinos will vary according to the velocity dispersion $<v^2>$ \cite{Freese83, Ruffini86}. The key point is that, for a gas of non-relativistic particles, $\rho_\nu$ and $<v^2>$ will depend on $T_\nu$ and $\xi_\nu$ in different ways, so that measuring independently the absorption factor and the free streaming length, it would be possible at least in principle to obtain the values of $T_\nu$ and $\xi_\nu$ without any ambiguity left.

What we have just said is even more true with respect to the LSS data, since even neutrinos with mass greater than $10^{-3}$~eV are in their non-relativistic regime during the late stages of the process of structure formation. We conclude then by stressing that one should be careful when parameterizing the lepton asymmetry by means of an effective number of degrees of freedom.

\section{Method}\label{sec:Method}
We used the \verb+CMBFAST+ code \cite{Se96}, modified as described in Ref. \cite{Le99} in order to account for a non vanishing chemical potential of neutrinos, to compute the temperature (TT) and polarization (TE) CMB spectra for different combinations of the cosmological parameters.
As a first step, we added three more parameters, namely the effective number of additional relativistic species $\DNoth$, the neutrino degeneracy parameter $\xi$ (both defined in Sec. \ref{sec:Basic formulae}) and the neutrino physical energy density $\omega_\nu\equiv\Omega_\nu h^2$ to the standard six-parameters $\conc$ model that accounts in a remarkably good way for the WMAP data. As anticipated above, $\DNoth$ accounts only for the extra energy density due to the presence of additional relic relativistic particles other than the three Standard Model neutrinos. We shall refer to the $(\omega_\nu,\, \xi,\,\DNoth)$ subspace as the ``neutrino sector'' of the parameter space (although, as we have just noticed, $\DNoth$ does not refer directly to neutrinos).

With the above mentioned choice of the parameters we can make a consistency check to our results, by verifying that imposing the priors $\xi = 0$, $\omega_\nu=0$ and $\DNoth=0$, we obtain results that are in agreement with the ones of the WMAP collaboration. Moreover, by choosing a sufficiently wide range for the variation of the three additional parameters, we can check how much their introduction affects the estimation of the best-fit values of the core parameters. Thus we choose to use the following parameters: the physical baryon density $\omega_b \equiv \Omega_b h^2$, the total density of non relativistic matter $\omega_m\equiv(\Omega_b+\Omega_\mathrm{CDM})h^2$, the scalar spectral index $n$, the optical depth to reionization $\tau$, the overall normalization of the CMB spectrum $A$, the physical neutrino density $\omega_\nu =\Omega_\nu h^2$, the neutrino degeneracy parameter $\xi$ and the extra energy density in non-standard relics $\DNoth$. We will be considering the scenario in which the three standard model neutrinos have all the same mass and chemical potential. We take the chemical potential to be positive (this corresponds to an excess of neutrinos over antineutrinos), but since the effects on the CMB do not depend on the sign of $\xi$, we quote the limits that we obtain in terms of its absolute value.  We do not include as a free parameter the Hubble constant $H_0$, whose degeneracy with the effective number of relativistic degrees of freedom and with the neutrino mass has been studied in previous works \cite{El03}. Instead we decided, according to the recent measurements of Hubble Space Telescope (HST) Key Project \cite{Fr01}, to assume that $h = 0.72$. Moreover, we restrict ourselves to the case of a flat Universe, so that the density parameter of the cosmological constant $\Omega_\Lambda$ is equal to $1-(\omega_m + \omega_\nu)/h^2$. We are thus dealing with a 8-dimensional parameter space.

Let us discuss in a bit more detail the way we deal with priors in the neutrino sector of parameter space, i.e., with information coming from other observations, and in particular from BBN. As we have stressed in section \ref{sec:Motivation}, the standard BBN scenario, together with the equalization of chemical potentials, constraints the neutrino degeneracy parameter to values lower than the ones considered in this paper; on the other hand this conclusion possibly does not hold in non-standard scenarios where additional relativistic relics are present. However, even non-standard scenarios of this kind usually single out some preferred region in parameter space. At the present, several non-standard scenarios that can account for the observed Helium abundance exist (see for example Refs. \cite{DiBari02} and \cite{DiBari03}) so that we adopt a conservative approach, and choose not to impose any prior on the neutrino sector, other than the ones that emerge ``naturally'' as a consequence of our choice of parameters. Anyway, this does not preclude the possibility of successively using the BBN information: in fact once the likelihood function in the neutrino sector has been calculated, it can be convolved with the relevant priors coming from non-standard BBN scenarios. 

We span the following region in parameter space: $0.020\leq\omega_b\leq0.028$,   $0.10\leq\omega_m \leq 0.18$, $0.9\leq n \leq 1.10$, $0 \leq \tau \leq 0.3$, $0.70\leq A \leq 1.10$, $0\leq\omega_\nu\leq0.30$, $0 \leq |\xi| \leq 2.0$, $0\leq\DNoth\leq 2.0 $. We shall call this our ``(5+3) parameter space''. In order to obtain the likelihood function ${\cal L}(\params)$ in this region, we sample it over a grid consisting of 5 equally spaced points in each dimension. For each point on our grid, corresponding to a combination of the parameters, we compute the likelihood relative to the TT \cite{WMAP:Hinshaw} and TE \cite{WMAP:Kogut} angular power spectrum observed by WMAP, using the software developed by the WMAP collaboration \cite{WMAP:methodology} and kindly made publicly available at their website\footnote{\url{http://lambda.gsfc.nasa.gov/}}.
To obtain the likelihood function for a single parameter, we should marginalize over the remaining ones. However for simplicity we approximate the multi-dimensional integration required for the marginalization procedure with a maximization of the likelihood, as it is a common usage in this kind of likelihood analysis. This approximation relies on the fact that the likelihood for cosmological parameters is supposed to have a gaussian shape (at least in the vicinity of its maximum) and that integration and maximization are known to be equivalent for a multivariate Gaussian distribution. 

According to Bayes' theorem, in order to interpret the likelihood functions as probability densities, they to be inverted through a convolution with the relevant priors, representing our knowledge and assumptions on the parameters we want to constrain. Here we shall assume uniform priors, i.e. we will assume that all values of the parameters are equally probable.

For each of the core parameters, we quote the maximum likelihood value (which we shall refer to also as the ``best-fit'' value) over the grid and the expectation value over the marginalized distribution function. We quote also the best chi square value $\chi_0^2$ (we recall that $\chi^2\equiv-2\ln{\cal L}$) divided by the number of degrees freedom, that is equal to the number of data (for WMAP, this is 1348) minus the number of parameters. For what concerns the parameters of the neutrino sector, we quote the maximum likelihood  and the expectation value as well, and in addition we report a $2\sigma$ confidence interval. Using a Bayesian approach, we define the 95\% confidence limits as the values at which the marginalized likelihood is equal to $\exp[-(\chi_0^2-4)/2]$, i.e., the values at which the likelihood is reduced by a factor $\exp(2)$ with respect to its maximum value~\footnote{The 95\% confidence level defined in this way is not in general equal to the $2\sigma$ region, defined computing the variance of the probability distribution. However, the two are equal for a gaussian probability density. As we shall see, almost all the marginalized distribution have a nearly gaussian shape. When it is not so, we shall point this out.}.
There is one exception to this procedure, namely, when the maximum likelihood value for a parameter that is positively defined (such as $\omega_\nu$ or the absolute value of $\xi$), let us call it $\theta$, is equal to zero. In this case, instead than computing the expectation value, we just give an upper bound. In order to do this, we compute the cumulative distribution function ${\cal C}(\theta)=\left(\int_0^{\theta}{\cal L}(\bar\theta) d\bar\theta\right) / \left(\int_0^\infty{\cal L}(\bar\theta) d\bar\theta\right)$ and quote as the upper limit at the 95\% confidence level the value of $\theta$ at which ${\cal C}(\theta) = 0.95$. 

Once we have obtained constraints on $\omega_\nu$, $\xi$ and $\DNoth$, we translate them to limits on the neutrino mass $m_\nu$, the lepton asymmetry $L$ and the extra number of effective relativistic species $\Delta\Neff$, using eqs. (\ref{eq:DNtot}) and (\ref{eq:DNnu}) together with the following relations {\cite{Freese83, Ruffini86, La03}}:
\begin{eqnarray}
&&\Omega_\nu h^2 = \sum_\nu \frac{m_\nu F(\xi_\nu)}{93.5\,\mathrm{eV}},\\[0.2 cm]
&&L\equiv \sum_\nu \frac{n_\nu-n_{\bar \nu}}{n_\gamma}= \nonumber\\[0.2cm]
&&=\frac{1}{12\zeta(3)}\left[\sum_\nu\left(\xi^3+\pi^2 \xi\right)\left(\frac{T^0_\nu}{T^0_\gamma}\right)^3\right],
\end{eqnarray}
where
\begin{multline}
F(\xi)\equiv\frac{2}{3\zeta(3)}\left[ \sum_{k=1}^\infty (-1)^{k+1}\frac{e^{+ k \xi}+e^{-k \xi}}{k^3} \right] = \\
= \frac{1}{3\zeta(3)}\left[\frac13\,\xi^3+\frac{\pi^2}{3}\xi
+4\sum_{k=1}^\infty(-1)^{k+1}\,\displaystyle\frac{e^{-k\xi}}{k^3}\right].
\end{multline}

\section{Results and discussion}\label{sec:Results}

We start our analysis by looking at the effect of the introduction of the additional parameters to the estimation of the core parameters $(\coreparams)$.
First of all, we check that imposing the priors $\xi = 0$, $\omega_\nu = 0$ and $\DNoth=0$ our results are in good agreement with the ones of the WMAP collaboration (we should refer to the values quoted in Table~I of Ref. \cite{WMAP:parameters}). The mean and maximum likelihood values that we obtain for each parameter are summarized in the second and third column of Table \ref{tab:core_summary}. We see that in all cases our results lie within the 68\% confidence interval of WMAP expected values. Then we remove the prior on $\omega_\nu$, while still retaining the ones on $\xi$ and $\DNoth$. The maximum likelihood model has still $\omega_\nu=0$.  The best-fit values of the core parameters are left unchanged, and the same happens for the best-fit $\chi^2$, thus suggesting that a non-zero $\omega_\nu$ is not required in order to improve the goodness of fit. The results for the core parameters are summarized in the fourth and fifth column of Table~\ref{tab:core_summary}.

\begin{center}
\begin{figure}[t!]
\resizebox{0.5\textwidth}{!}{\includegraphics{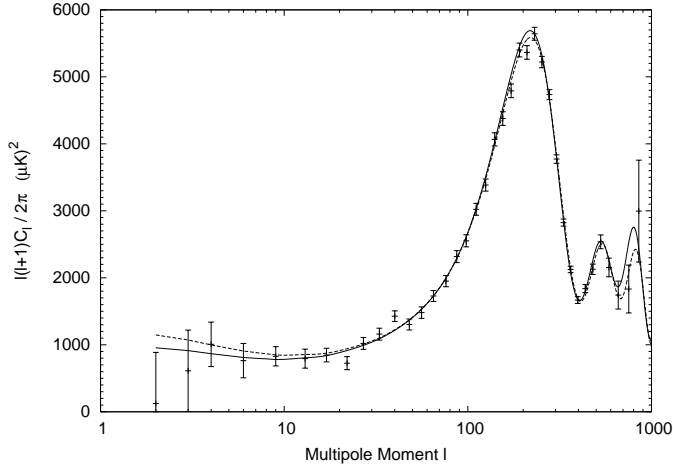}}
\caption{Comparison between the best-fit power spectrum obtained assuming the priors $\xi=0, \DNoth=0$ (solid line) and without such prior (dashed line). The points are the WMAP data on the temperature angular power spectrum \cite{WMAP:Hinshaw}.}
\label{fig:compare_spectra}
\end{figure}
\end{center}
\begin{center}
\begin{figure*}[t!]
\resizebox{\textwidth}{!}{
\includegraphics{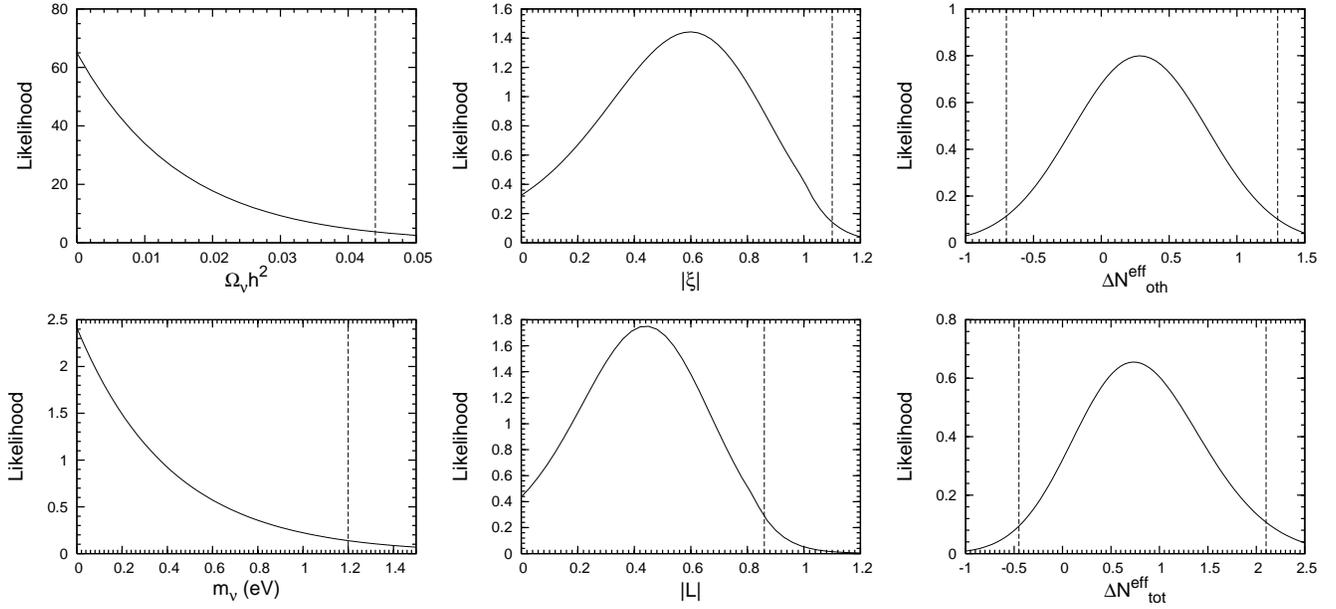}
}
\caption{Likelihood functions for $\omega_\nu$, $|\xi|$, $\DNoth$ and for the derived parameters $|L|$, $m_\nu$ and $\Delta\Neff$, obtained as a result of the analysis in (5+3) parameter space. The dotted lines bound the 95\% confidence interval. The functions are normalized so that their integral is equal to unity.}
\label{fig:6Like_5+3}
\end{figure*}
\end{center}

\begingroup
\squeezetable
\begin{table*}
\begin{ruledtabular}
\caption{Core Parameters}
\label{tab:core_summary}
\begin{tabular}{ldddddd}
&  \multicolumn{2}{c}{Priors: $\xi =0$, $\Omega_\nu=0$, $\Delta\Neff_{\mathrm{others}}=0$}& \multicolumn{2}{c}{Priors: $\xi =0$, $\Delta\Neff_{\mathrm{others}}=0$} & \multicolumn{2}{c}{No Neutrino Priors} \\
 & \multicolumn{1}{c}{} & \multicolumn{1}{c}{Maximum} & \multicolumn{1}{c}{} & \multicolumn{1}{c}{Maximum} & \multicolumn{1}{c}{} & \multicolumn{1}{c}{Maximum} \\
Parameter & \multicolumn{1}{c}{Mean} & \multicolumn{1}{c}{Likelihood} & \multicolumn{1}{c}{Mean} & \multicolumn{1}{c}{Likelihood} & \multicolumn{1}{c}{Mean} & \multicolumn{1}{c}{Likelihood} \\
\hline\\
Baryon density,$\omega_b$      & 0.024 & 0.024 & 0.024 & 0.024 & 0.023 & 0.022\\
Matter density,$\omega_m$      & 0.15  & 0.16 & 0.15  & 0.16 & 0.14  & 0.14\\
Hubble constant\footnote{The value of the Hubble constant is kept fixed to $h=0.72$.},$h$            & 0.72 & 0.72  & 0.72 & 0.72  & 0.72 & 0.72 \\
Spectral index,$n$             & 1.00 & 1.00 & 1.00 & 1.00 & 0.98 & 0.95\\
Optical depth,$\tau$           & 0.13 & 0.075 & 0.13 & 0.075 & 0.12 & 0.075\\
Amplitude,$A$								   & 0.8 &  0.8 & 0.8 &  0.8 & 0.8 &  0.7\\
$\chi^2/\nu$		   &  & \multicolumn{1}{c}{1437/1343} & & \multicolumn{1}{c}{1437/1342} & & \multicolumn{1}{c}{1431/1340}
\end{tabular}
\end{ruledtabular}
\end{table*}
\endgroup
Finally, we compute the likelihood over our whole parameter space. The results for the core parameters are summarized in the sixth and seventh column of Table \ref{tab:core_summary}. The maximum likelihood model over the grid has $(\params)=(0.022,\,0.14,\,0.95,\,0.075,\,0.7,\,0,\,0.5,\,0)$. We see that this time, the best fit values for the five core parameters are slightly changed with respect to the standard case. The changes in $\omega_m$ and $n$ could seem strange at a first sight, since intuitively one would expect the opposite behaviour, i.e., a change to larger values for both, because a larger $\omega_m$ could keep the time of matter-radiation equality, while a larger $n$ would increase the power on small scales thus leaving more room for neutrino free streaming. This is because the goodness of fit of a particular model with respect of the WMAP data is mainly determined by its ability to fit the first and second peak. Increasing together $\omega_m$, $n$, $\xi$ and $\DNoth$ would increase the height of the first peak that can be then lowered back by decreasing the overall amplitude $A$. We show in fig. \ref{fig:compare_spectra} a comparison between the best-fit spectrum in the $(\xi = 0,\,\DNoth=0)$ subspace with the best-fit spectrum on the whole space.

Now let us turn our attention to the neutrino sector of parameter space.
The best-fit model over the $(\xi=0,\,\DNoth=0)$ subspace of the grid has $\omega_\nu=0$, and the $\chi^2$ changes from 1437 to 1541 when going from $\omega_\nu = 0$ to the next value in our grid, $\omega_\nu = 0.075$. We can compute an upper bound for $\omega_\nu$, but
since the region in which ${\cal L}(\omega_\nu)$ significantly differs from zero is all comprised between the first two values in our grid $\{0, 0.075\}$, the result is rather dependent from the particular interpolation scheme we choose. Using a simple, first order interpolation scheme, we find the bound $\omega_\nu < 0.0045$ (95\% CL), corresponding to $m_\nu<0.14$~eV, while using higher order interpolation schemes the bound weakens up to $\omega_\nu < 0.015$ ($m_\nu<0.47$~eV). This result should then be taken with caution and we shall simply consider it as an indication that, although we are using a grid-based method with a rather wide grid spacing instead than the more sophisticated Markov Chain Monte Carlo (MCMC) method \cite{Chris01,Lewis02,book:Gamerman}, we basically obtain the same results of the WMAP collaboration, namely, $\omega_\nu\le0.0072$ \cite{WMAP:parameters}, when imposing the priors $\xi =0,\,\DNoth=0$.

We make a second check by imposing that $\omega_\nu=0,\,\xi=0$ and computing the 95\% confidence region for $\DNoth$. We find that $0\le\DNoth\le1.4$. Since the degeneracy parameter is vanishing, the same limit applies to $\Delta\Neff=\DNoth$. This is quite in agreement with the results quoted in Ref. \cite{Cr03}, although it is more restrictive. This is probably due to the fact that we are imposing a stronger prior on $h$, keeping it constant and equal to 0.72. This is confirmed by a visual inspection of fig. 2 of Ref. \cite{Cr03}.

The best-fit value for neutrino density over the whole parameter space is still $\omega_\nu=0$, but this time $\chi^2$ changes from 1431 to 1441 as $\omega_\nu$ goes from 0 to 0.075, so that the probability density spreads out to higher values of $\omega_\nu$ with respect to the preceding case. The result is that the upper bound raises up to $\omega_\nu<0.044$, quite independently from the interpolation scheme used. This is probably related to the already observed trend for which, when the energy density of relativistic relics is increased, the possibility for larger neutrino masses reopens \cite{Ha03, El03, Ha04, Lesg01}. 

The maximum likelihood value for the degeneracy parameter is $|\xi| = 0.5$, while the expectation value over the distribution function is $|\xi| = 0.56$ (corresponding to $|L|=0.43$). At the $2\sigma$ level, the degeneracy parameter is constrained in the range $0\le|\xi|\le1.07$. This corresponds to $0\le |L| \le 0.9$. For what concerns the additional number of relativistic relics, the maximum likelihood model has $\DNoth=0$, and the expectation value over the marginalized probability function is $\DNoth=0.3$. The 95\% confidence region is $-0.7\le\DNoth\le1.3$. This opening towards smaller, negative values of $\DNoth$ can be ascribed to the fact that such values produce a lowering of the acoustic peak, that can be compensated by a larger degeneracy parameter. The quoted bounds on $\omega_\nu$ and $\xi$ translate to the following bound on the neutrino mass: $m_\nu<1.2$~eV (95\% CL). In fig. \ref{fig:6Like_5+3} we show the likelihood functions, while in Table \ref{tab:nu_summary} we summarize our results for the basic and derived parameters describing the neutrino sector.

We remark that, although the maximum likelihood model over the whole grid has $\omega_\nu=0$, this is not in contradiction with our choice of considering $\xi$ and $\DNoth$ as independent parameters, in spite of the fact that in this limit they should be degenerate. The basic reason is that, as can be seen from the likelihood curves, models with $\omega_\nu>0$ can be statistically significant. For these models, $\Delta\Neff_\nu$ and $\DNoth$ are not exactly degenerate.

In order to better study the partial degeneracy between $|\xi|$ and $\DNoth$, and then to understand how the value of $\DNoth$ affects the estimation of the degeneracy parameter, we compute the likelihood curve for the degeneracy parameter for particular values of $\DNoth$. The results are shown in Table \ref{tab:xi_vs_DNoth}. . From this table, a quite evident trend appears, namely that for large values of $\DNoth$, smaller values of $|\xi|$ are preferred, and viceversa. As already noticed, this is probably related to the fact that when $\DNoth$ is increased, it remains less room for the extra energy density of neutrinos coming from the non-vanishing degeneracy parameter. It is worth noting that, for $\DNoth\simeq 0$, the case $\xi=0$ lies outside the 95\% confidence region. We stress the fact that, according to theoretical predictions, in models of degenerate BBN with ``3+1'' neutrino mixing, if chemical potentials are large ($\xi>0.05$), the production of sterile neutrinos is suppressed, effectively resulting in $\DNoth=0$ \cite{DiBari02, DiBari03}.

\begingroup
\squeezetable
\begin{table}
\begin{ruledtabular}
\caption{Neutrino Sector}
\label{tab:nu_summary}
\begin{tabular}{ld}
 &  \multicolumn{1}{c}{95\% Confidence }   \\
Parameter & \multicolumn{1}{c}{Interval} \\
\hline\\
Physical neutrino density, $\omega_\nu$  & \lesssim 0.044 			 \\[0.1cm]
Degeneracy parameter, $|\xi|$            & 0.60^{+0.50}_{-0.60}  \\[0.1cm]
Neutrino mass in eV, $m_\nu$             & \lesssim 1.2          \\[0.1cm]
Lepton asymmetry, $|L|$                  & 0.46^{+0.43}_{-0.46}  \\[0.1cm]
Effective number of additional           & 0.30\pm1.0 \\
relativistic relics,  $\DNoth$           &              \\[0.1cm]
Effective number of additional           & 0.70^{+1.40}_{-1.15} \\
relativistic relics,  
$\Delta N^{\mathrm{eff}}$                & 

\end{tabular}
\end{ruledtabular}
\end{table}
\endgroup

\begingroup
\squeezetable
\begin{table}
\begin{ruledtabular}
\caption{Correlation between $\xi$ and $\DNoth$}
\label{tab:xi_vs_DNoth}
\begin{tabular}{cd}
$\DNoth$    & \multicolumn{1}{c}{$\xi$}   \\
						& \multicolumn{1}{c}{(95\% Confidence Interval)} \\
\hline\\      
  0      & 0.65\pm 0.58          \\[0.1cm]
  0.5    & 0.42^{+0.58}_{-0.42}  \\[0.1cm]
  1.0    & 0.18^{+0.58}_{-0.18}  \\[0.1cm]
  1.5    & \le 0.53          \\[0.1cm]
  2.0    & \le 0.29            
\end{tabular}
\end{ruledtabular}
\end{table}
\endgroup

\section{Conclusions and perspectives}
\label{sec:Conclusions}
In this paper, we have studied the possibility to constraint the lepton asymmetry of the Universe, the sum of neutrino masses, and the energy density of relativistic particles using the WMAP data, in the framework of an extended flat $\conc$ model. Despite the fact that the current amount of cosmological data can be rather coherently explained by the standard picture with three thermally distributed neutrinos, vanishing lepton asymmetry and no additional particle species, nevertheless we think that it is useful to explore how non-standard scenarios are constrained by the cosmological observables. We have concentrated our attention to models with a (eventually large) net lepton asymmetry (corresponding to a non-zero degeneracy parameter for neutrinos). Such models are motivated in the framework of extensions to the standard model of particle physics, and can possibly explain the observed amount of baryon asymmetry in the Universe. Having in mind this, we have also included the energy density of relativistic species as an independent parameter. In this last aspect, our approach differs from previous ones, where the two parameters where considered	degenerate. We have remarked that, although an approximate degeneracy between the two exists, it could be broken by finite mass effect, especially in the case of neutrino masses saturating the tritium beta decay bound.

  When considering perfect lepton symmetry , our results are in agreement with previous ones. In the more general case, we have found that, at the $2\sigma$ level the bounds on the degeneracy parameter and lepton asymmetry are respectively $0\le |\xi| \le 1.1$ and $0\le |L| \le 0.9$. The effective number of additional relativistic species (excluding the contribution from the non standard thermal distribution of neutrinos) is bounded as follows (95\% CL): $-0.7~\le~\DNoth~\le~1.3$. Including also neutrinos, this reads $-0.45\le\Delta\Neff\le2.10$. This limit is much more restrictive than the ones found in similar analysis \cite{Cr03, Pi03}. This is probably due to the fact that we assume a very strong prior on the Hubble parameter, fixing $h=0.72$. The physical explanation is that the later matter-radiation equality due to $\Delta\Neff>0$ can be compensated by making $\omega_m=\Omega_mh^2$ larger, and viceversa. This gives rise to a partial degeneracy between $\Delta\Neff$ and $h$, thus making the costraints on both parameters looser unless some external prior is imposed to break the degeneracy. 
  
  We also find that the data are compatible with $\omega_\nu$ and $m_\nu$ equal to $0$, with upper bounds (95\% CL) $\omega_\nu\le 0.044$ and $m_\nu\le 1.2\,\eV$. This bounds are larger than the ones usually found, and this is probably due on one hand to the presence of a larger energy density of UR particles, and on the other hand to the wide grid spacing we have used.
  
  The usual scenario, with $|L|=0$ and $\Delta\Neff=0$, is then compatible with WMAP data at the $2\sigma$ level; however the likelihood curves show that alternative scenarios with $\xi\simeq 0.6$ and $\Delta\Neff\simeq 0.7$ have a larger likelihood with respect to the data. In effect, the standard scenario lies outside the $1\sigma$ confidence region. Even if this is not enough to definitely claim evidence, in the CMB anisotropy spectrum, of exotic physics, we think that it is however interesting that non-standard models are not ruled out but actually preferred by the WMAP data.
  
  We have also studied how the results on the lepton asymmetry can change when more precise information on the energy density of relativistic particles is given. We have shown that, the smaller is the extra energy density, the larger is the allowed lepton asymmetry. In particular, for models with vanishing $\DNoth$, perfect lepton symmetry is ruled out at the $2\sigma$ level. This is probably due to the approximate degeneracy between $\Delta \Neff$ and $\xi$. The issue of the exact extent of this degeneracy is still open, and we think that it deserves a deeper attention. It would be desirable to investigate if future precision CMB experiments, and in particular the PLANCK mission, can clearly disentangle the two parameters. Even more promising to this purpose is the power spectrum of LSS, as we have stressed in Sec. \ref{sec:Effect}.
  
 The results presented in this paper have been derived assuming that the three neutrino families have all the same mass and chemical potential, owing to the structure of the \verb+CMBFast+ code used to generate the theoretical power spectra. It is yet to investigate the role of a non-uniform distribution of the lepton asymmetry between different families. In a similar way we did not investigate models in which chemical equilibrium between neutrinos and antineutrinos does not hold, implying $\xi_\nu \neq - \xi_{\bar \nu}$.
 
We conclude then that WMAP data still cannot exclude the presence of non-standard physics in the early evolution of the Universe. In particular, they do not exclude the presence of a large neutrino asymmetry, and consequently they do not rule out exotic leptogenesis scenario where a large lepton number is produced. 

\begin{acknowledgments}
ML would like to thank Giovanni Montani for useful discussion and comments.
\end{acknowledgments}

\end{document}